\documentclass{iopart}

\usepackage{iopams}  
\usepackage{graphicx}  

\begin{document}

\vspace*{-2.cm}

\title[NLO analysis of jet production in Pb+Pb collisions at the LHC]
{NLO analysis of inclusive jet, tagged jet and di-jet production in Pb+Pb collisions at the LHC}

\author{I. Vitev$^\S$}

\address{$^\S$Los Alamos National Laboratory, 
Theoretical Divisions, Mail Stop B283 \\ Los Alamos, NM 87545, USA \\ }


\begin{abstract}
We present results and predictions at next-to-leading  order for the recent LHC lead-lead run at a center-of-mass 
energy of 2.76~TeV per  nucleon-nucleon pair. Specifically, we focus on the suppression the 
single and double inclusive jet cross sections  and demonstrate how the di-jet asymmetry, recently
measured by ATLAS and CMS, can be extracted from this general result. 
The case of jets tagged by an electroweak boson is exemplified by the $Z^0$+jet channel. We predict 
a signature  transition from enhancement to suppression of the tagged jet related to the medium-induced 
modification of the parton shower.  Finally, we clarify the relation between the suppression of 
inclusive jets, tagged jets and di-jets and the quenching 
of inclusive particles on the example of the recent ALICE charged hadron attenuation data. 
\end{abstract}

\vspace*{-.7cm}

\section{Introduction}

Jets physics is a new area of active research at RHIC and at the LHC that paves the way for novel 
tests of QCD multi-parton  dynamics in heavy-ion reactions~\cite{Vitev:2008rz}. At present, 
perturbative QCD calculations  of hard probes in “elementary” nucleon-nucleon  reactions can be 
consistently combined with the effects  of the nuclear medium up to next-to-leading order~\cite{jets}. 
While such accuracy  is desirable for leading particle tomography, it is absolutely essential for 
the new jet observables.  With this motivation, we present results and predictions at NLO for the 
recent LHC lead-lead run at a center-of-mass energy of  2.76 TeV per nucleon-nucleon 
pair~\cite{qmtalks}. Our analysis includes not only final-state inelastic parton 
interactions in the QGP, but also initial-state cold nuclear matter effects and non-perturbative 
hadronization corrections.

\section{Jets production and attenuation in heavy ion reactions}

The basis for the evaluation of multi-jet cross sections in heavy ion collisions are
the corresponding cross sections in the more elementary nucleon-nucleon reactions.
Accurate perturbative calculations at next-to-leading order 
are essential to capture the cross section dependence of the jet radius 
$R=\sqrt{\delta \eta^2 + \delta \phi^2}$~\cite{jets} 
and provide an accurate estimate of the transverse energy $E_T$ dependence of
tagged jets~\cite{Neufeld:2010fj} and di-jets~\cite{He:2011pd}. For inclusive jets and 
di-jets we use the EKS NLO results~\cite{EKS} and obtain baseline cross sections in
excellent agreement with experiment from RHIC to LHC energies. For $Z^0$-tagged jets
we use the MCFM code that works very well at the Tevatron and at the 
LHC~\cite{Campbell:2003dd}.

\subsection{Quenching of inclusive jets}

In Ref.~\cite{Ovanesyan:2011xy} it was shown that in QCD the final-state
process-dependent medium-induced radiative corrections factorize in the production cross sections
on the example of a singe jet. The generalization to multiple jets proceeds as follows.
\begin{eqnarray}
\!\!\!\!\! d \sigma(\epsilon_1,\cdots,\epsilon_n)^{n-{\rm jet}}_{\rm quench.} &=&
d\sigma(\epsilon_1,\cdots,\epsilon_n)^{n-{\rm jet}}_{\rm pp}  \otimes P_1(\epsilon_1)
\cdots  \nonumber \\
\!\!\!\!\! && \otimes P_n(\epsilon_n)
\, \ |J_1(\epsilon_1)| \cdots |J_n(\epsilon_n)| \;. \qquad
\label{fact}
\end{eqnarray}
Here, $P_i(\epsilon_i)$ is the probability that the $i^{\rm th}$ jet will lose a fraction
$\epsilon_i$ of its energy, $|J_i(\epsilon_i)|$ are phase space Jacobians, and
$\otimes$ denotes an integral convolution. Any dependence on the jet
reconstruction parameters is not shown explicitly in Eq.~(\ref{fact}).

\begin{figure}[!t]
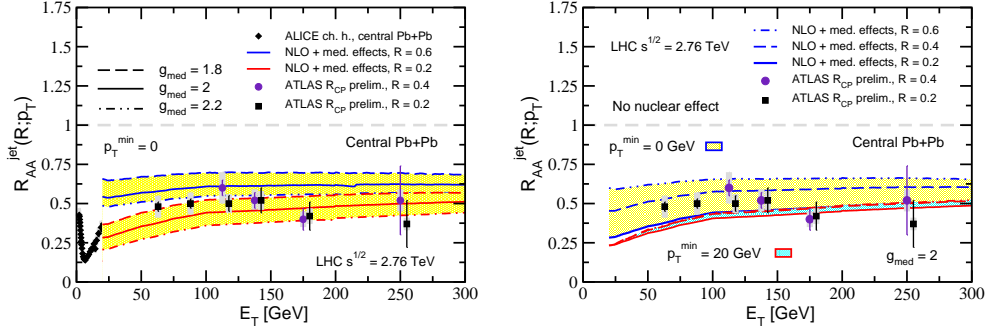

\begin{center}
\vspace*{+.6cm}
\includegraphics[width=6.3cm]{LHCcouplingQM.eps}
\hspace*{.2cm} \includegraphics[width=6.3cm]{PTnoPTQM.eps}
\end{center}
\vspace*{-0.2cm}
\caption{ Left panel: predicted suppression of single inclusive jets in central Pb+Pb
collisions at the LHC for two different radii $R=0.2, 0.6$ and three couplings between the jet and the 
medium $g_{\rm med} = 1.8, \, 2,\, 2.2$. Preliminary ATLAS data for $R=0.2, 0.4$ is also shown.
Right panel: jet radius dependence of the inclusive quenching factor with ($p_T^{\rm min}=20$~GeV)
and without ($p_T^{\rm min}=0$~GeV) diffusion of the parton shower energy due to collisional
processes.} 
\label{figure1}
\vspace*{-2mm}
\end{figure} 

Theoretical predictions for the attenuation of inclusive jets are shown in the 
left panel of Figure~\ref{figure1} for two radii $R=0.2, 0.6$. Our results include
final-state QGP-induced radiative energy loss effects~\cite{Vitev:2007ve} and 
initial-state cold nuclear matter effects~\cite{Vitev:2008vk,Neufeld:2010dz}. 
Preliminary ATLAS results~\cite{qmtalks} are also shown for comparison. 
Collisional interactions of the in-medium parton shower can re-distribute a 
fraction the jet energy to the non-Abelian plasma away from the jet 
axis~\cite{source}. This is simulated in the right panel of Figure~\ref{figure1} 
by a  parameter $p_T^{\rm min}=20$~GeV. The main result is that such diffusion 
eliminates any residual radius $R$ dependence of the inclusive jet quenching~\cite{He:2011pd}. 

\subsection{$Z^0$-tagged jets}

\begin{figure}[!t]
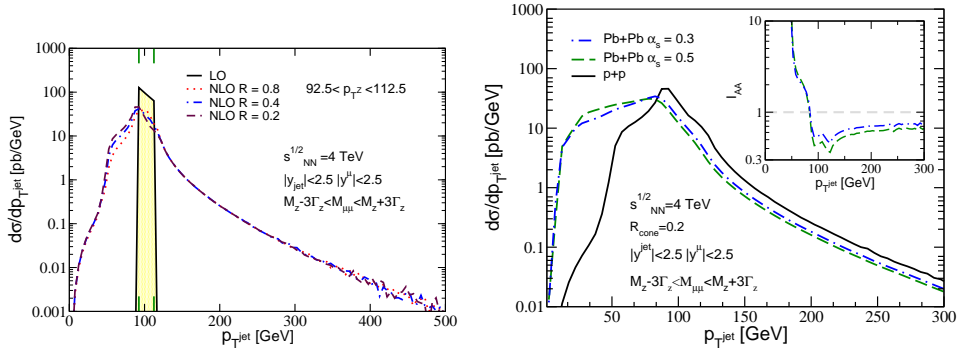

\vspace*{.cm}
\begin{center}
\includegraphics[width=6.cm]{925_1125_I.eps}
\hspace*{.1cm}
\includegraphics[width=6.3cm]{925_1125_r02.eps}
\end{center}
\vspace*{-0.cm}
\caption{Left panel:  comparison of LO and NLO jet cross sections for jets tagged by a 
$Z^0 \rightarrow \ell^++\ell^-$ at the LHC. 
Right panel: QGP-induced nuclear modification of $Z^0$-tagged jets in central Pb+Pb reactions 
at the LHC. Insert shows the predicted 
transition from moderate quenching for $p_T^{\rm jet} >  p_T^{Z^0}$ to  strong 
enhancement  $p_T^{\rm jet} <  p_T^{Z^0}$. }
\label{figure3}
\vspace{-2mm}
\end{figure}

$Z^0$-tagged jets have been proposed as one of the important new channels that
open up at the LHC to study parton energy loss and jet quenching. To utilize its
potential, NLO calculations are necessary~\cite{Campbell:2003dd}. The left panel of 
Figure~\ref{figure3} shows that at tree level the jet $p_T^{\rm jet}$ coincides with the momentum 
of the  $Z^0$, here measured via its di-lepton decay channel. ${\cal O}(G_F\alpha_s^2)$ 
is the first non-trivial order at which $Z^0$-tagged jets can be studied~\cite{Neufeld:2010fj}. 
The right panel of Figure~\ref{figure3} shows the predicted QGP-induced modification for
jets of radius $R=0.2$ associated with a $Z^0$ tag in central Pb+Pb collisions. The insert
shows a characteristic transition from  suppression  ($I_{AA}< 1$) for $p_T^{\rm jet} > p_T^{Z^0}$ 
to strong enhancement ($I_{AA} \gg 1$) for $p_T^{\rm jet} < p_T^{Z^0}$. Integrating out the
jet and including the relevant tree level processes we evaluate the inclusive $Z^0$ production 
cross section. We find that cold nuclear matter effects~\cite{Vitev:2008vk,Neufeld:2010dz} are 
small and the calculation accurately predicts the  $Z^0$ cross section measured in Pb+Pb
collisions measured by CMS~\cite{Chatrchyan:2011ua}.

\subsection{Di-jets and their $E_T$ asymmetry}

\begin{figure}[!t]
\vspace*{+1.cm}
\begin{center}
\includegraphics[width=6.5cm]{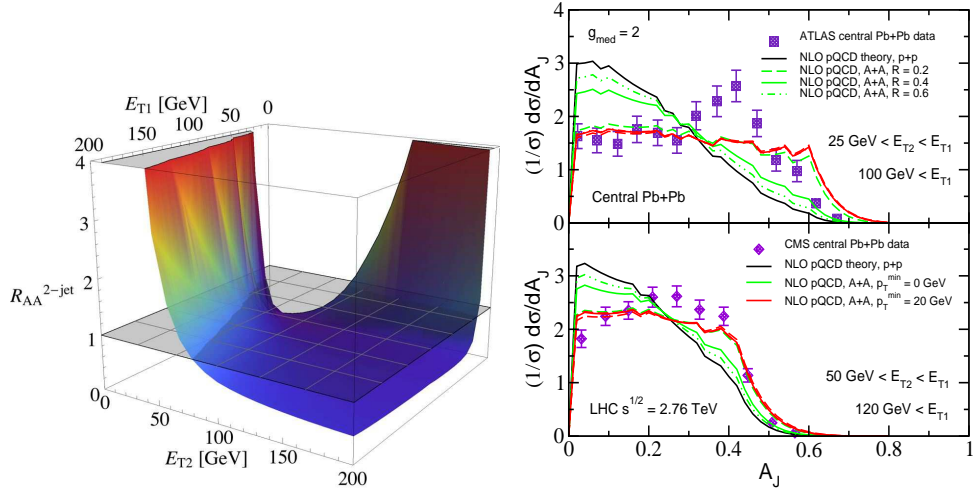}
\hspace*{.2cm}\includegraphics[width=6.cm,height=6.5cm]{AJnewAArad.eps}
\end{center}
\vspace*{-.2cm}
\caption{ Left panel: nuclear modification for di-jets in central Pb+Pb reactions
at the LHC. Both jet radii are chosen to be $R=0.2$.  
Right panel: the corresponding enhanced di-jet asymmetry is compared to ATLAS and CMS
experimental data for $R=0.2,\, 0.4,\ 0.6$. Green lines include radiative energy loss
and red lines, denoted $p_T^{\rm min} = 20$~GeV, add the diffusion of the parton shower
energy through collisional processes. } 
\label{figure2}
\vspace{-2mm}
\end{figure}

New insights into the medium modification of parton showers are provided by
the 2D attenuation pattern of di-jets. A specific subset of di-jets in the general
$(E_{T_1},E_{T_2})$ plane are the ones of fixed energy asymmetry 
$A_J = (E_{T_1} - E_{T_2})/( E_{T_1} - E_{T_2})$. We have verified that 
NLO calculations of di-jets reproduce with excellent accuracy the baseline
asymmetry measured in p+p collisions by the ATLAS experiment. The left
panel of Figure~\ref{figure2} shows the first calculation for the suppression
of di-jets $R_{AA}^{\rm 2-jet}$ in central Pb+Pb reactions at the LHC~\cite{He:2011pd}. It is 
characterized by a broad region of approximately constant suppression
around $ E_{T_1}= E_{T_2} $ and strong enhancement for $ E_{T_1} \ll E_{T_2} $,
$ E_{T_1} \gg E_{T_2} $. The corresponding QGP-enhanced asymmetry is shown in the
the right panel of Figure~\ref{figure2} for jet radii $R=0.2,\, 0.4,\, 0.6$.
We find that radiative energy loss~\cite{Vitev:2007ve} (green lines) can 
explain approximately 1/2 of the measured $A_J$ broadening~\cite{Aad:2010bu,Chatrchyan:2011sx}.
Diffusion of the parton shower energy away from the jet axis through collisional
processes~\cite{source} eliminates the residual jet radius dependence and the theoretical 
calculation is compatible with the experimental measurements for all~$R$.

\section{Leading particle suppression}

\begin{figure}[!t]
\vspace*{+.2cm}
\begin{center}
\includegraphics[width=6.3cm,height=5.5cm]{PTfeedALICE.eps}
\hspace*{.2cm}\includegraphics[width=6.3cm,height=6.5cm]{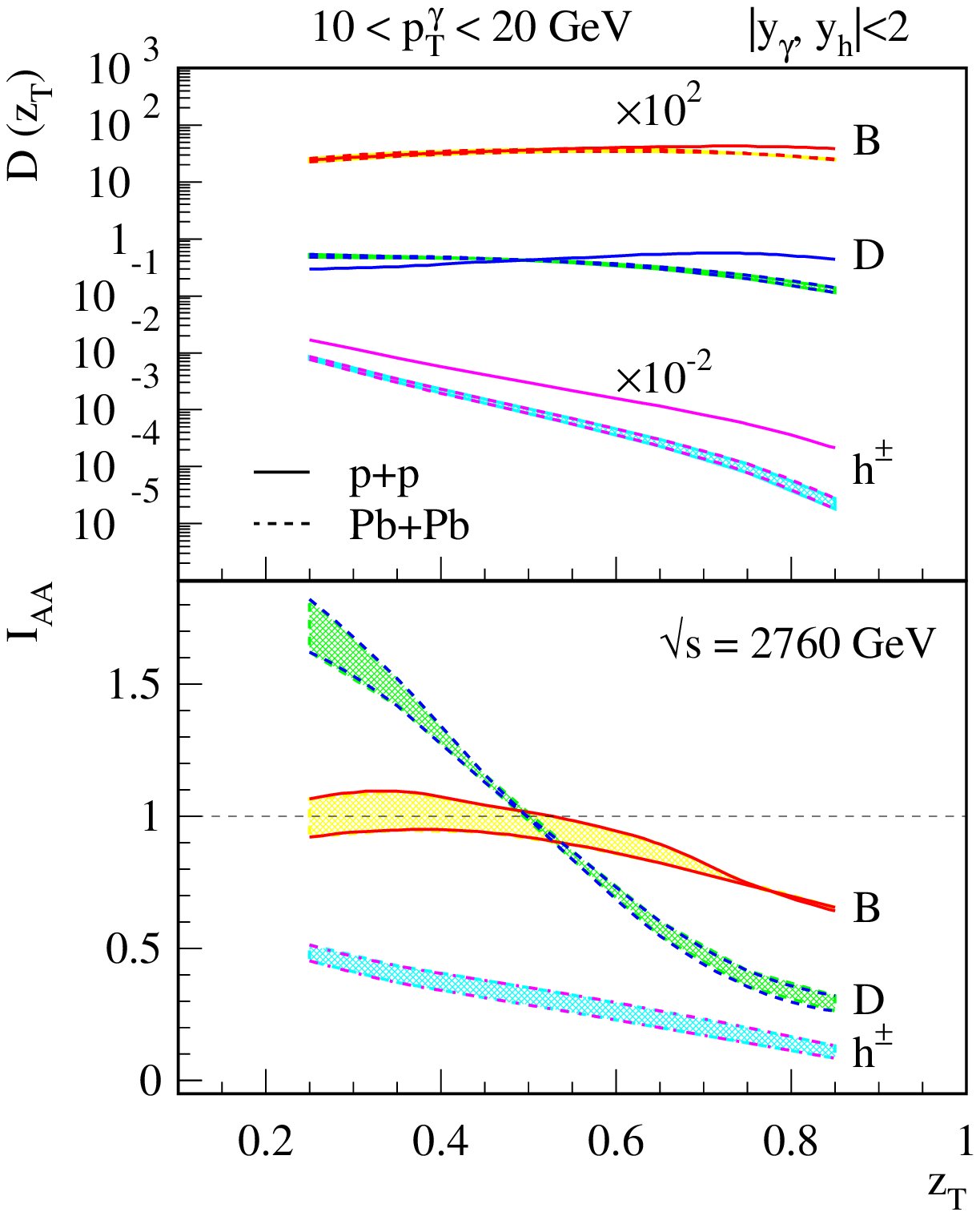}
\end{center}
\vspace*{-0.3cm}
\caption{Left panel: comparison of predictions for inclusive pion 
quenching with (solid lines) and without (dashed lines) gluon feedback to recent ALICE data
in central Pb+Pb reactions at the LHC. Right panel: predicted effective fragmentation 
functions in photon-triggered light and heavy hadron production in central Pb+Pb reactions 
at the LHC. We also show the difference in the nuclear modification $I_{AA}$ for light and
heavy flavor, to be tested in the future. }
\label{figure5}
\vspace{-2mm}
\end{figure}

The quenching of jets and hadrons has to be understood in the
same theoretical formalism. Specifically, the nuclear modification of jets for
small radii approximates the suppression of leading particles~\cite{Vitev:2008rz}.  
An example of an early theoretical prediction~\cite{Vitev:2005he} is given in the left
panel of Figure~\ref{figure5} and compared to ALICE  data on charged hadron
attenuation in central Pb+Pb collisions~\cite{Aamodt:2010jd}. The contribution of the 
medium-induced bremsstrahlung to hadron production below $p_T = 6$~GeV is significant and
is reflected in the non-trivial and non-monotonic $p_T$ dependence of $R_{AA}(p_T)$, confirmed by 
experiment. Cold nuclear matter effects also play a role in this intermediate $p_T$ 
region. Further constraints on the mechanisms of jet quenching can be obtained 
by investigating open heavy flavor production~\cite{hf}. Preliminary results are already
available at the LHC~\cite{Dainese:2011vb}. It will be important and illuminating in the
future to extend these measurements to photon-tagged heavy meson production. Theoretical
predictions for the differences in the effective fragmentation functions between light
and heavy flavor $D(z_T)$ and the nuclear modification factor $I_{AA}(z_T)$ are presented 
in the right panel of Figure~\ref{figure5}.

\section{Summary}

We presented  NLO results and predictions for inclusive jet, $Z^0$-tagged jet, and di-jet production in 
heavy ion collisions at the LHC.  We also demonstrated the relation between the quenching of jets, 
the attenuation of leading particles, and the modification of photon-tagged light and heavy 
hadron distributions. We found that in all cases there is a good qualitative understanding    
of the suppression of hard probes in Pb+Pb collisions at the LHC. Experimental results have 
already provided guidance on the directions where further theoretical developments are desirable for
improved quantitative description of heavy ion reactions at the high energy frontier. Namely,
these are the interaction between the parton showers and the medium and the energy loss mechanisms 
for heavy quarks.

\vspace*{0.2cm}\noindent\textbf{Acknowledgment:} I would like to thank my collaborators
Y.~He, Z.B.~Kang, R.B.~Neufeld, G.~Ovanesyan, R.~Sharma and B.W. Zhang for helpful discussion.
This work is supported by the U.S. Department of Energy  Office of Science 
under contract No.  DE-AC52-06NA25396.

\end{document}